\documentclass[fleqn,usenatbib]{mnras}

\usepackage{newtxtext,newtxmath}

\usepackage[T1]{fontenc}

\usepackage{subfig}

\DeclareRobustCommand{\VAN}[3]{#2}
\let\VANthebibliography\thebibliography
\def\thebibliography{\DeclareRobustCommand{\VAN}[3]{##3}\VANthebibliography}

\usepackage{graphicx}	
\usepackage{amsmath}	

\title[A lensed FRB candidate]{A lensed FRB candidate in the first CHIME/FRB Catalogue and its potential implications}

\author[C M. Chang et al.]{
Chenming Chang$^{1,2}$\thanks{E-mail: changcm@pmo.ac.cn},
Songbo Zhang$^{1,3}$\thanks{E-mail: sbzhang@pmo.ac.cn},
Di Xiao$^{1,2}$\thanks{E-mail: dxiao@pmo.ac.cn},
Zhenfan Tang$^{1,2}$,
Ye Li$^{1,2}$,
Junjie Wei$^{1,2}$,
\newauthor and Xuefeng Wu$^{1,2}$\thanks{E-mail: xfwu@pmo.ac.cn}
\\
$^{1}$Purple Mountain Observatory, Chinese Academy of Sciences,\\Nanjing 210023, China\\
$^{2}$School of Astronomy and Space Sciences, University of Science and Technology of China,\\Hefei 230026, China\\
$^{3}$ CSIRO Space and Astronomy, Australia Telescope National Facility, PO Box 76, Epping, NSW 1710, Australia\\
}

\date{Accepted XXX. Received YYY; in original form ZZZ}

\pubyear{\the\year{}}

\begin{document}
\label{firstpage}
\pagerange{\pageref{firstpage}--\pageref{lastpage}}
\maketitle

\begin{abstract}
Fast radio bursts (FRBs) are immensely energetic radio pulses with durations of milliseconds. Given their high all-sky rate, the probability of an FRB being lensed by an intervening massive object is non-negligible. In this study, we search for possible lensing candidates within the first Canadian Hydrogen Intensity Mapping Experiment FRB catalogue using an autocorrelation algorithm and verification through signal simulations. We identify FRB 20190308C as a lensed candidate with a significance of 3.4$\sigma$. Furthermore, we constrain the mass of the lensing object using the Chang–Refsdal lens model, based on the flux ratio and time delay between the substructures of FRB 20190308C. Future long-term and high-precision observations are expected to reveal more lensed FRBs.
\end{abstract}




\section{Introduction} \label{sec:introduction}

Fast radio bursts (FRBs) are bright millisecond-long radio flashes first discovered in 2007 \citep{Lorimer2007}. Even before confirming their cosmological origin in 2017 \citep{Chatterjee2017, Tendulkar2017}, FRBs were proposed as excellent cosmological probes \citep[e.g.,][]{Deng2014}. While most currently observed FRBs have a single-pulse morphology, a small fraction have been detected with multiple bursts. Due to the limited sensitivity and survey strategies of radio telescopes, some single-pulse FRBs may be intrinsically repeating, with only one burst captured. Although repetition might be an intrinsic property of FRBs, propagation effects along their travel path could also cause variations in their pulse profiles. For instance, time delays caused by gravitational lensing can split millisecond-scale FRB signals into multiple bursts.

According to modern theories of gravity, gravitational lensing occurs when light rays travel along geodesics and are deflected due to the intervening inhomogeneities in the matter distribution. The most typical case of lensing involves quasars being lensed by foreground galaxies or galaxy clusters. It is because quasars are widely distributed in the universe and are very bright, making them visible from large distances \citep{2022ChPhL..39k9801L}. These lensed systems have been suggested as powerful probes to study several astrophysical issues, such as the spatial mass distribution at kiloparsec and subkiloparsec scales, the overall geometry, kinematics and components of the universe, and distant objects that are too small or too faint to be resolved or detected with current instruments \citep{2010ARA&A..48...87T, 2019RPPh...82l6901O}. In addition to quasars, FRBs, as transient explosive sources, with tens of thousands of signals that will be guaranteed in the future, have gained attention as potential targets for lensing studies, given that a small part of them are expected to be lensed and detected.

Strong gravitational lensing can produce multiple bursts from a single burst, as strongly lensed FRBs would present multiple images \citep{2021ApJ...912..134C, 2021ApJ...923..117C}. Although the angular separation of these images is usually too small to be resolved by current telescopes, they can still be identified through their different arrival times. \citet{2016PhRvL.117i1301M} proposed that compact objects, such as massive compact halo objects (MACHOs) or primordial black holes (PBHs), could cause a single-pulse FRB to display a multi-peaked profile. \citet{2018A&A...614A..50W}, \citet{2020ApJ...896L..11L} and \citet{2020ApJ...900..122S} also suggested that lensed FRBs could be used to constrain the fraction of compact objects in dark matter. \citet{2020ApJ...896L..11L} attempted to identify lensed signals using dynamic spectra, while \citet{2020ApJ...900..122S} discussed the potential for resolving temporal structures on the order of $\sim 10 \;ns$, considering that FRBs might have much smaller substructures. However, neither study found credible lensed signals. Similarly, \citet{2022ApJ...928..124Z} used a cross-correlation method to analyse the latest $\sim 600$ FRBs but did not identify any lensed FRBs. \citet{2022PhRvD.105j3528K} employed an autocorrelation method on a sample of 143 FRBs, finding two possible candidates. \citet{2022PhRvD.106d3017L} used a novel interferometric method to search for lensed FRBs with a time resolution of $1.25 \;ns$, corresponding to very tiny lenses, but found no lenses in 172 bursts. 

Notably, most of these works focused on the classical point-mass model, which suggests a brighter leading peak. 
\citet{2022ApJ...928..124Z} considering those FRBs with brighter trailing peaks, but still did not find any lensed signals.
In this work, we employed the autocorrelation method to search for potential lensed FRBs in the first CHIME/FRB catalogue \citep{2021AAS...23832501M}. We used the Chang–Refsdal lens which allows for a brighter trailing peak, to constrain the lens mass following the method proposed by \citet{2021ApJ...912..134C}. Additionally, we use the time-dependent dispersion ($\sigma$) as a threshold, which differs from the fixed constant threshold used in \citet{2022PhRvD.105j3528K}.

This paper is organized as follows. Section \ref{sec:search} presents the method we adopt to identify lensing signatures in FRBs. In section \ref{sec:model}, we introduce the Chang–Refsdal lens and constrain the lens mass based on observed properties, including the time delay and flux ratio of the lensed FRB candidate we identified. Conclusions and discussions are provided in section \ref{sec:conclusion}.

\section{Search for potential lensing candidates} \label{sec:search}

\subsection{Method} \label{sec:method}

As suggested by \citet{2021ApJ...918L..34W}, signal autocorrelation can be used to measure the time delay of signals of a gravitationally lensed system. We define the autocorrelation function (ACF) of the light curve $I_t$ as: 
\begin{equation}
    C(k)=\frac{\sum_{t=0}^{N-k}(I_t-\bar{I})(I_{t+k}-\bar{I})}{\sum_{t=0}^N(I_t-\bar{I})^2},
\end{equation}
where $\bar{I}$ represents the mean value of $I_t$. According to \citet{2018PhRvD..98l3523J}, a lensed signal will exhibit a spike at $\delta t = \Delta t$, where $\Delta t$ represent the time delay between the lensed peaks. For real data, the ACF typically shows a spiky structure,  meaning $\delta t = \Delta t$ may not be the only peak. Therefore, we need to define a threshold to distinguish lensing-induced spikes from noise. Here we apply Gaussian smoothing to the ACF, choosing $1/3 dt$ as the standard deviation of the one-dimensional Gaussian kernel to minimize the influence of adjacent points on the central point, where $dt = 0.98\,\rm ms$ is the time resolution of the data. 
The dispersion between the ACF and the smoothed function $F(k)$, following \citet{2021ApJ...918L..34W}, is defined as:
\begin{equation}
    \sigma^2=\frac{1}{N}\sum_{j=0}^N[C(k)-F(k)]^2,
\end{equation}
where $N$ is the total number of bins. A $3\sigma$ threshold is used to distinguish lensed systems from noise.

\subsection{Data}
The data used in this study is from the first CHIME/FRB catalogue \citep{2021AAS...23832501M}, which contains 536 FRBs. 
The parameters produced in this catalogue indicate that 42 of these FRBs exhibit multi-peak structures, suggesting that some could be potential candidates for lensed bursts. 
Due to the existence of noise in the raw data, the result from the autocorrelation method could be significantly impacted. Therefore, we also applied this method to the model-fitted data provided in the catalogue. As long as either the original data or the model-fitted data of a signal passes the screening, we would add it to the list of potential candidates.

We processed the data from the 16,384 channels available in the CHIME catalogue by averaging every 512 channels, resulting in 32 frequency channels. We then calculated the time series for each of these channels. 
The channels containing signals were selected, their time series concatenated, and the ACF for the total time series was computed. 
Since lensed signals maintain similar characteristics across different frequency channels, we assume that this process does not affect the ACF's calculation for lensed signals. 
However, for signals that vary across different frequency channels, this approach may weaken the ACF's results.

\begin{figure}
  \centering
  \subfloat[Waterfall]{
    \label{fig.wfall}
    \includegraphics[width=0.75\linewidth]{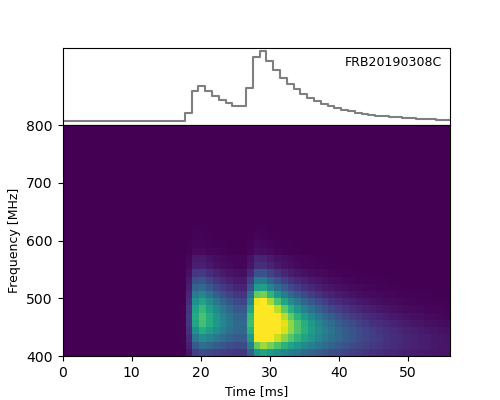}
  }
  
  \subfloat[Residual]{
    \label{fig.residual}
    \includegraphics[width=0.75\linewidth]{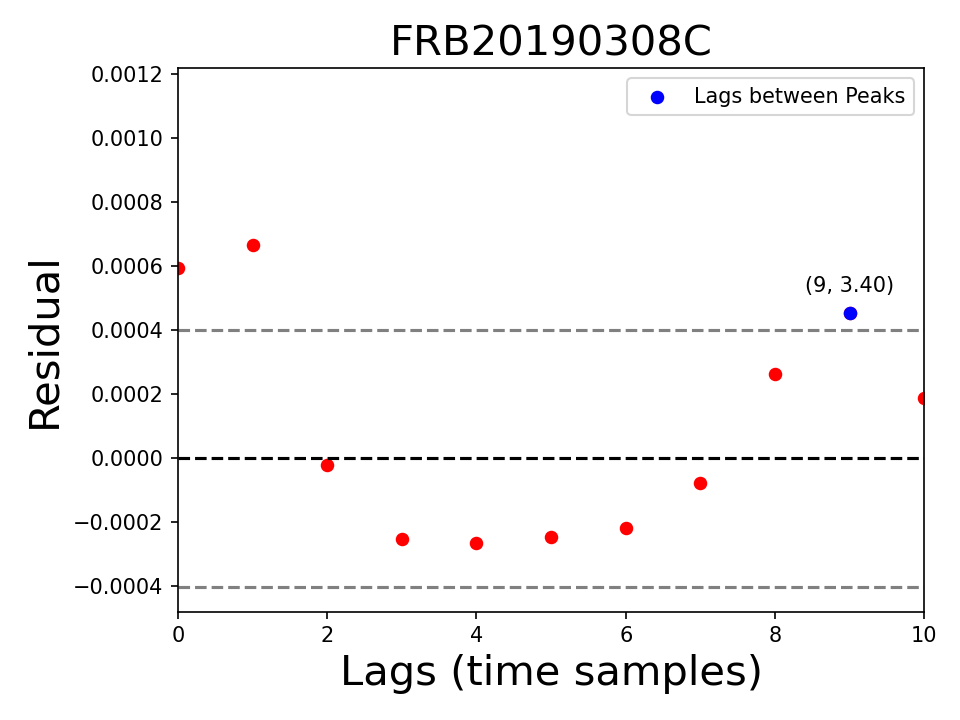}
  }
  \caption{The model-fitted waterfall of FRB 20190308C (top panel) and the residual between the autocorrelation and its Gaussian smoothing (bottom panel). Since the residuals have a peak value $3.40\sigma$ when the time delay is $9$ 9 times the time resolution (the blue dot), we believe the two substructures of FRB 20190308C are similar.}
  \label{fig.wfall & fig.residual}
\end{figure}

Figure \ref{fig.wfall & fig.residual} shows 1 shows an example of the waterfall plot of the model-fitted data for FRB 20190308C (top panel) and the corresponding residuals between the ACF of the time series and its Gaussian smoothing in units of $\sigma$ defined in section \ref{sec:method} (bottom panel). Since only the significance of the autocorrelation between the two possible lensed substructures is relevant, we display only the cut of the autocorrelation from delay $= 0$ to delay $= d+w$ in the bottom panel, where $d$ is the time delay between the two peaks, and $w$ represents the width of one peak based on time samples. The black dashed line represents the Gaussian smoothing of the ACF, the gray dashed line indicates the $3\sigma$ range, the red dots represent the residuals between the ACF and the Gaussian smoothing at the corresponding time delays, and the blue dots indicate the time delay between the two peaks of the signal. The number in the parentheses indicates a signal significance of $3.40\sigma$, which is higher than the $3\sigma$ threshold at a time delay of $9 dt$. This suggests that the two substructures have similar time series in each channel.

To address the possibility that model-fitted data might lose fine temporal structure present in the original data, we calculated the residuals between the model-fitted and the original data. To investigate the impact of this on the search, we used the same method to calculate the autocorrelation function of the residuals, to search for temporal structures that might have not been captured by the model fit. The results indicate that there is no significant correlation in the residual autocorrelation for these 42 signals. 
If the background noise is sufficiently clean, correlated fine structures might still be present in the residuals after model smoothing of a lensed signal. Therefore, a null search in the residual auto-correlation might indicate a non-detection of lensing. However, due to the complex nature of background noise components, analysing the signal solely through its dynamic spectrum is challenging. Hence, this paper focuses specifically on the lensing possibility of the model-fitted pulsing signals.


It is also worth noting that this method only considers changes in flux as a function of time. 
To confirm a signal as a lensed candidate, we need to analyse the dynamic spectrum of each peak. This requires using the dynamic spectrum, as suggested by \citet{2020ApJ...896L..11L}.

\subsection{Result}
After applying the autocorrelation method to the 42 FRBs with multi-peaks provided in the first CHIME/FRB catalogue, we identified a total of 10 bursts that satisfied the filter criteria. These bursts are: FRB 20180917A, FRB 20181019A, FRB 20181028A, FRB 20181222A, FRB 20190122C, FRB 20190208A, FRB 20190308C, FRB 20190601C, FRB 20190605B, and FRB 20190625E.

Among these 10 bursts, only one, FRB 20181028A, met the filter criteria for both its original and model-fitted data. The original data of FRB 20181222A, FRB 20190122C and FRB 20190208A passed the inspection, but their model-fitted data did not meet the criteria. Conversely, for the remaining bursts, although their original data did not pass the filter criteria, their model-fitted data yielded positive results after calculating the ACF.

\begin{figure}
    \centering
    \includegraphics[width=1\linewidth]{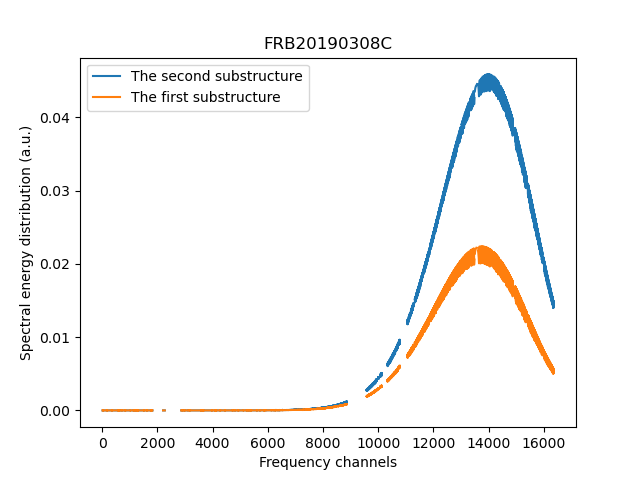}
    \caption{The spectra of the substructures of FRB 20190308C. We calculate the cross-correlation of the two spectra and the cross-correlation function reaches its maximum value when the offset is zero which implies that the spectra of the two substructures are similar.}
    \label{fig:pinpu}
\end{figure}

As suggested by \citet{2016PhRvL.117i1301M} and \citet{2020ApJ...896L..11L}, lensed pulses should exhibit differences in flux magnification and time delay, while maintaining almost identical profiles. 
Therefore, it is essential to compare the distribution of substructures of each signal in terms of frequency.
We calculated the cross-correlation function between the spectra of different sub-structures of each signal. The consistency of the spectral distribution of the sub-structures was assessed based on the peak position of the cross-correlation function and the distance between the spectral peaks of the substructures.  
Based on the cross-correlation function analysis of the substructures of each signal, only FRB 20190308C passed the screening criteria. Figure \ref{fig:pinpu} shows the spectra of the substructures of FRB 20190308C. The spectra of the two substructures are considered similar, as the cross-correlation function reaches its maximum value when the offset is zero.

\subsection{Simulation}

For signals that have passed the initial screening criteria, the existing conditions may still not be robust enough to confirm the possibility of lensing. 

According to \citet{2021AAS...23832501M}, the model-fitted data is treated as a superposition of different components. A signal consisting of $N$ components can be represented as follows:
\begin{equation}
    S = \sum_{i=0}^N A_i \times F_i \times T_i,
\end{equation}
where $A_i$ is the overall amplitude of the $i$th burst component; $F_i\left(\gamma_i,r_i\right)$ is the time-independent spectral energy distribution and $T_i\left(DM,t_{arr,i},w_i,\tau\right)$ is the temporal shape of the burst. Although we attempted to identify similar components in the time series structure by calculating the autocorrelation function, there are slight differences in various parameters of these similar components, such as signal width ($w_i$) and spectral indices ($\gamma_i$ and $r_i$) of the spectral energy distribution since these parameters were uniquely defined for each of the $N$ components. 

A lensed system theoretically requires these parameters to remain consistent across different substructures, while deviations always exist in practice. These differences may be inherent to the signal components themselves, or they may be caused by noise interference or fitting errors. We cannot distinguish how these differences arise, but we can simulate lensed FRB signals by controlling these parameters to remain consistent within the substructures of the signal, in order to investigate whether lensing effects can produce similar dynamic spectra of the candidates.

To strengthen our results, we also employ signal simulation. 
By controlling the parameters of the substructure to be consistent and simulating results that are most similar to the observed signals, we can determine whether these signals could plausibly originate from lensed systems.

The simulation software \textsc{Simfred} \citep{2023MNRAS.523.5109Q} was used to simulate the lensing signals. 
We set the parameters of the substructures to be identical, with only the flux magnification and arrival time differing. 
Using the parameters of the first substructure of FRB 20190308C, we performed our simulation and ensured that the parameters of the second substructure, such as signal width and spectrum, were consistent with those of the first substructure to generate the simulated signal. 
We then concatenated the time series of the simulated signal with that of the model-fitted signal to create a total time series. 
Their similarity was calculated using the autocorrelation method described in Section  \ref{sec:method}.

\begin{figure}
  \centering
  \subfloat[Waterfall]{
    \label{fig.wfall_2}
    \includegraphics[width=0.75\linewidth]{FRB20190308C_wfall.png}
  }

  \subfloat[Simulation]{
    \label{fig.simulation}
    \includegraphics[width=0.75\linewidth]{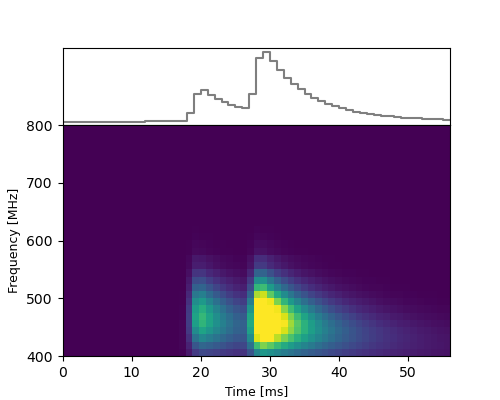}
  }

  \subfloat[Residual]{
    \label{fig.Simulateautocorrelation}
    \includegraphics[width=0.75\linewidth]{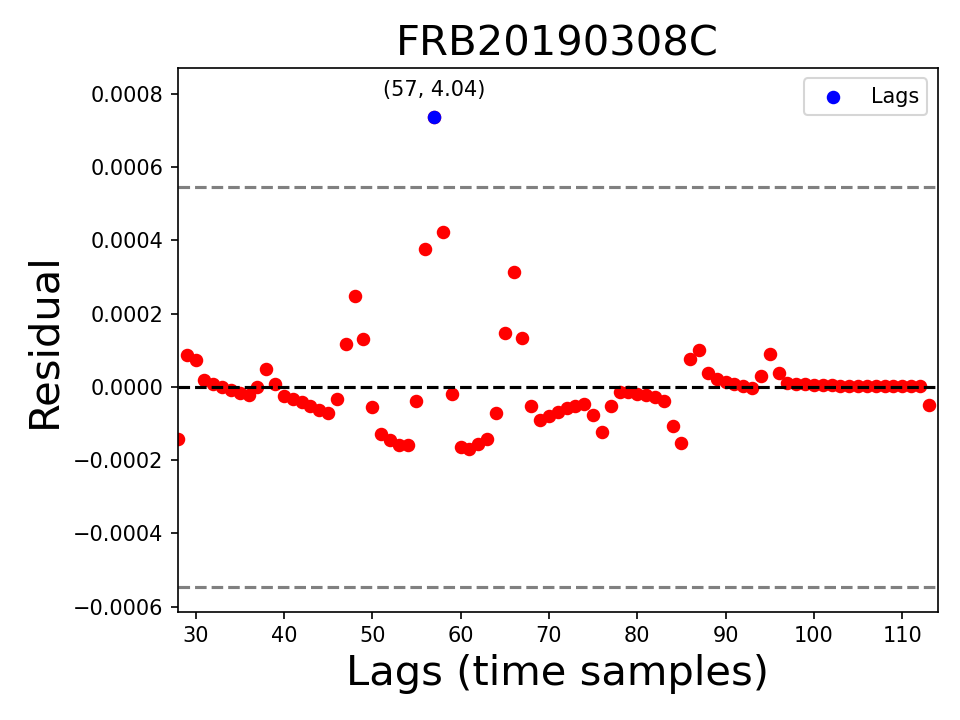}
  }
  \caption{The model-fitted waterfall of FRB 20190308C (top panel), the simulation obtained using Simfred(middle panel) and the residual between the autocorrelation of the total time series and its Gaussian smoothing (bottom panel).}
  \label{fig.wfall_2 & fig.simulation}
\end{figure}

Figure \ref{fig.wfall_2 & fig.simulation} shows the model-fitted waterfall (top panel), the simulation of FRB 20190308C (middle panel), and the residual between the autocorrelation of the total time series and its Gaussian smoothing (bottom panel). The result we obtained is a significance of 4.04 $\sigma$, which exceeds the 3$\sigma$ threshold. Therefore, we consider the two signals to be similar, suggesting that FRB 20190308C could originate from a lensed system.

\section{Physical Model of Gravitational Lensing} \label{sec:model}
\subsection{Model introduction}
Although FRB 20190308C meets our filter criteria and its pulses exhibit similar profiles in the dynamic spectrum, it cannot be explained by the classical point-mass lens model. The point-mass lens model predicts a fainter trailing pulse, whereas the trailing peak of FRB 20190308C is brighter. 
However, it is still possible to explain this feature using the Chang–Refsdal lens model \citep{1979Natur.282..561C,2006MNRAS.369..317A,2021ApJ...912..134C}, suggesting that FRB 20190308C could be produced by a gravitationally lensed system.

The Chang-Refsdal lens model considers a star as a point-mass lens with the galaxy providing a background perturbation field. According to \citet{1979Natur.282..561C}, \citet{2006MNRAS.369..317A}, \citet{2021ApJ...912..134C} and \citet{2022MNRAS.516..453G}, the lens equation in dimensionless form is given by:
\begin{equation}
    \begin{aligned}
        y_1 &= (1+\gamma)x_1-\frac{x_1}{x_1^2+x_2^2},\\
        y_2 &= (1-\gamma)x_2-\frac{x_2}{x_1^2+x_2^2},
    \end{aligned}
\end{equation}
where $(x_1,x_2)$ and $(y_1,y_2)$ represent positions on the defector plane and the observer plane, respectively. 
Although the equation can be solved in principle, the actual algebra is quite complicated. 
Therefore, applying a simple numerical algorithm to solve the quartic polynomial is more advantageous. Fortunately, if the source lies along either of the symmetry axes of the system, the complete analytic solutions are relatively simple to derive.

Since the number of unknowns exceeds the number of equations, the lens mass cannot be directly obtained through the Chang-Refsdal lens model.
\citet{2021ApJ...912..134C} proposed a method to constrain the mass using the flux ratio and time delay between the substructures of signals.

For $y_1 = 0$, the time delay and flux ratio can be written as 
\begin{equation}\label{t_y1=0}
    \Delta t = \frac{4GM}{c^3}\left(1+z_l\right)\left[\frac{y_2s_2}{2\left(1-\gamma\right)}+\ln\left(\frac{s_2+y_2}{s_2-y_2}\right)\right]
\end{equation} 
and 
\begin{equation}\label{R_y1=0}
    R = \frac{\left(y_2^2+2+y_2s_2\right)\left(y_2s_2-4\gamma\right)+8\gamma^2+2\gamma y_2^2}{\left(y_2^2+2-y_2s_2\right)\left(y_2s_2+4\gamma\right)-8\gamma^2-2\gamma y_2^2},
\end{equation} 
where $\gamma$ is the external shear strength, $z_l$ is the redshift of the lens and
\begin{equation}
    s_2 = \sqrt{y_2^2+4\left(1-\gamma\right)}.
\end{equation}
The corresponding formulas for $y_2 = 0$ are: 
\begin{equation}\label{t_y2=0}
    \Delta t = \frac{4GM}{c^3}\left(1+z_l\right)\left[\frac{y_1s_1}{2\left(1+\gamma\right)}+\ln\left(\frac{s_1+y_1}{s_1-y_1}\right)\right]
\end{equation} 
and 
\begin{equation}\label{R_y2=0}
    R = \frac{\left(y_1^2+2+y_1s_1\right)\left(y_1s_1+4\gamma\right)+8\gamma^2-2\gamma y_1^2}{\left(y_1^2+2-y_1s_1\right)\left(y_1s_1-4\gamma\right)-8\gamma^2+2\gamma y_1^2},
\end{equation} 
where
\begin{equation}
    s_1 = \sqrt{y_1^2+4\left(1+\gamma\right)}.
\end{equation}
When the source is located at the tips of the inner caustics, the minimum time delay can be written as 
\begin{equation}\label{mint}
    \Delta t = \frac{4GM}{c^3}\left(1+z_l\right)\left[\frac{2\gamma}{1-\gamma^2}+\ln\left(\frac{1+\gamma}{1-\gamma}\right)\right].
\end{equation}

\subsection{Constraints on lens mass}
\begin{figure}
    \centering
    \includegraphics[width=1\linewidth]{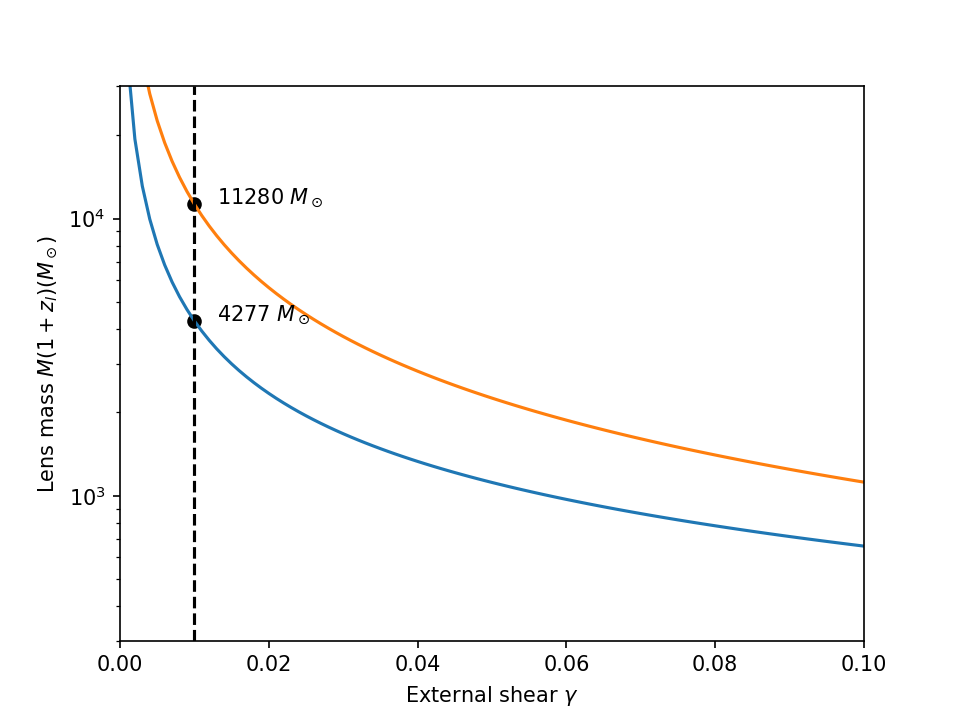}
    \caption{The range of lens mass constraints corresponding to the variation of $\gamma$ from 0 to 0.1. The blue is given by Equations (\ref{t_y1=0}) and (\ref{R_y1=0}), and the orange line is obtained by Equation (\ref{mint}). The black vertical dashed line represents the condition when $\gamma = 0.01$ and the combined lens mass range is $\left[4277 M_\odot, 11280 M_\odot\right]$}.
    \label{fig:mass range}
\end{figure}

As discussed in Section \ref{sec:model}, for FRB 20190308C, we can derive constraints on the lens mass using its flux ratio $R = 0.5$ and time delay $\Delta t = 8.85 \,\rm ms$, as shown in Figure \ref{fig:mass range} based on the chosen value of $\gamma$. Here we only consider the condition when $0< \gamma < 1$ following the results presented in \citet{2021ApJ...912..134C}. The blue line in the figure are obtained from Equations (\ref{t_y1=0}) and (\ref{R_y1=0}), and the orange line from Equation (\ref{mint}). It is clear that the range of the combined mass $M\left(1+z_l\right)$ varies with different values of $\gamma$.
The black dashed line taken as an example represents the lens mass constraints for $\gamma=0.01$. For $R = 0.5$, $\Delta t = 8.85\,\rm ms$ and $\gamma = 0.01$, the limit of the combined lens mass is $\left[4277 M_\odot, 11280 M_\odot\right]$.

\section{Discussion and Conclusions} \label{sec:conclusion}

In this work, we performed the autocorrelation method on data from the first CHIME/FRB catalogue to identify potential lensed FRBs. We find that the ACF calculated with raw data is easily affected by noise, thereby affecting the reliability of identifying similar structures in FRB light curves using ACF.
Therefore, we also used model-fitted data, significantly enhancing the S/N.
Out of the 42 FRBs exhibiting multiple peaks, our method identified 10 candidates using the ACF approach. 
Only one of these FRBs passed the $3 \sigma$ filter criteria with both raw and model-fitted data. Three met the screening criteria with their raw data alone, while the remaining six signals satisfied the filter criteria solely with their model-fitted data. 
Since substructures of lensed signals should have almost the same dynamic spectrum, we compared their spectra and ruled out nine FRBs due to inconsistencies. Finally, only FRB 20190308C was identified as a plausible candidate for gravitational lensing.

While the ACF indicates that the signal has lensing characteristics, it does not confirm that the origin is a lensed system. 
To strengthen our findings, we simulated the signal of FRB 20190308C by controlling the structural parameters of its substructures as required by the lens model.
We used the method described in Section \ref{sec:method} to compare the simulated signal with the original signal to demonstrate the possibility that the signal could be generated by a lensed system from another perspective. 
The ACF shows a significant peak of $4.04 \sigma$, suggesting that the two time series are similar, indicating that a lensed system could generate a signal with a profile like FRB 20190308C.

If a lensed signal shows a pulse profile similar to FRB 20190308C, where the first peak is weaker than the second one, this cannot be explained by the point-mass lens model.
Therefore, we used the Chang–Refsdal lens model and obtained a range of the combined lens mass as shown in Figure \ref{fig:mass range}. The mass range varies with different values of $\gamma$. If $0< \gamma <1$, for a chosen value of $\gamma = 0.01$, the combined lens mass can be constrained to be $\left[4277 M_\odot, 11280 M_\odot\right]$. The case when $\gamma >1$ is more complex and will be discussed in future work.

It is worth noting that the possibility of FRBs generating similar structures both temporally and spectrally through random processes always exists. Such coincidentally generated similar structures and those produced by gravitational lensing are almost indistinguishable based solely on the dynamic spectrum of the signal. Therefore, this paper seeks only potential lensed FRB candidates rather than definitive lensed FRBs. According to \citet{2022ChPhL..39k9801L}, to identify lensed FRB signals, it is necessary to find at least two pairs of bursts with the same time delay, flux ratio and DM difference, or to observe inconsistent source positions. Therefore, future observations with longer durations and higher precision will be required to find convincing lensed signals.
What we present in this article are only potential candidates for lensed FRB signals.

\section*{Acknowledgements}

This research has been partially funded by the National SKA Program of China (2022SKA0130100,2020SKA0120300), the National Natural Science Foundation of China (grant Nos. 12041306, 12273113,12233002,12003028,12321003), the CAS Project for Young Scientists in Basic Research (Grant No. YSBR-063), the International Partnership Program of Chinese Academy of Sciences for Grand Challenges (114332KYSB20210018), the National Key R\&D Program of China (2021YFA0718500), the ACAMAR Postdoctoral Fellow, China Postdoctoral Science Foundation (grant No. 2020M681758), and the Natural Science Foundation of Jiangsu Province (grant Nos. BK20210998).

\section*{Data Availability}

The data underlying this work are available in \citep{2021AAS...23832501M} and in its online supplementary material. The source code \textsc{Simfred} \citep{2023MNRAS.523.5109Q} to simulate FRBs for this work is publicly available at the following repository: \url{https://github.com/hqiu-nju/simfred/}.




\bibliographystyle{mnras}
\bibliography{FRB_lensing} 

\begin{thebibliography}{}
\makeatletter
\relax
\def\mn@urlcharsother{\let\do\@makeother \do\$\do\&\do\#\do\^\do\_\do\%\do\~}
\def\mn@doi{\begingroup\mn@urlcharsother \@ifnextchar [ {\mn@doi@} {\mn@doi@[]}}
\def\mn@doi@[#1]#2{\def\@tempa{#1}\ifx\@tempa\@empty \href {http://dx.doi.org/#2} {doi:#2}\else \href {http://dx.doi.org/#2} {#1}\fi \endgroup}
\def\mn@eprint#1#2{\mn@eprint@#1:#2::\@nil}
\def\mn@eprint@arXiv#1{\href {http://arxiv.org/abs/#1} {{\tt arXiv:#1}}}
\def\mn@eprint@dblp#1{\href {http://dblp.uni-trier.de/rec/bibtex/#1.xml} {dblp:#1}}
\def\mn@eprint@#1:#2:#3:#4\@nil{\def\@tempa {#1}\def\@tempb {#2}\def\@tempc {#3}\ifx \@tempc \@empty \let \@tempc \@tempb \let \@tempb \@tempa \fi \ifx \@tempb \@empty \def\@tempb {arXiv}\fi \@ifundefined {mn@eprint@\@tempb}{\@tempb:\@tempc}{\expandafter \expandafter \csname mn@eprint@\@tempb\endcsname \expandafter{\@tempc}}}

\bibitem[\protect\citeauthoryear{{An} \& {Evans}}{{An} \& {Evans}}{2006}]{2006MNRAS.369..317A}
{An} J.~H.,  {Evans} N.~W.,  2006, \mn@doi [\mnras] {10.1111/j.1365-2966.2006.10303.x}, \href {https://ui.adsabs.harvard.edu/abs/2006MNRAS.369..317A} {369, 317}

\bibitem[\protect\citeauthoryear{{Chang} \& {Refsdal}}{{Chang} \& {Refsdal}}{1979}]{1979Natur.282..561C}
{Chang} K.,  {Refsdal} S.,  1979, \mn@doi [\nat] {10.1038/282561a0}, \href {https://ui.adsabs.harvard.edu/abs/1979Natur.282..561C} {282, 561}

\bibitem[\protect\citeauthoryear{{Chatterjee} et~al.,}{{Chatterjee} et~al.}{2017}]{Chatterjee2017}
{Chatterjee} S.,  et~al., 2017, \mn@doi [\nat] {10.1038/nature20797}, \href {https://ui.adsabs.harvard.edu/abs/2017Natur.541...58C} {541, 58}

\bibitem[\protect\citeauthoryear{{Chen}, {Shu}, {Zheng}  \& {Li}}{{Chen} et~al.}{2021a}]{2021ApJ...912..134C}
{Chen} X.,  {Shu} Y.,  {Zheng} W.,   {Li} G.,  2021a, \mn@doi [\apj] {10.3847/1538-4357/abf119}, \href {https://ui.adsabs.harvard.edu/abs/2021ApJ...912..134C} {912, 134}

\bibitem[\protect\citeauthoryear{{Chen}, {Shu}, {Li}  \& {Zheng}}{{Chen} et~al.}{2021b}]{2021ApJ...923..117C}
{Chen} X.,  {Shu} Y.,  {Li} G.,   {Zheng} W.,  2021b, \mn@doi [\apj] {10.3847/1538-4357/ac2c76}, \href {https://ui.adsabs.harvard.edu/abs/2021ApJ...923..117C} {923, 117}

\bibitem[\protect\citeauthoryear{{Deng} \& {Zhang}}{{Deng} \& {Zhang}}{2014}]{Deng2014}
{Deng} W.,  {Zhang} B.,  2014, \mn@doi [\apjl] {10.1088/2041-8205/783/2/L35}, \href {https://ui.adsabs.harvard.edu/abs/2014ApJ...783L..35D} {783, L35}

\bibitem[\protect\citeauthoryear{{Gao} et~al.,}{{Gao} et~al.}{2022}]{2022MNRAS.516..453G}
{Gao} H.-X.,  et~al., 2022, \mn@doi [\mnras] {10.1093/mnras/stac2215}, \href {https://ui.adsabs.harvard.edu/abs/2022MNRAS.516..453G} {516, 453}

\bibitem[\protect\citeauthoryear{{Ji}, {Kovetz}  \& {Kamionkowski}}{{Ji} et~al.}{2018}]{2018PhRvD..98l3523J}
{Ji} L.,  {Kovetz} E.~D.,   {Kamionkowski} M.,  2018, \mn@doi [\prd] {10.1103/PhysRevD.98.123523}, \href {https://ui.adsabs.harvard.edu/abs/2018PhRvD..98l3523J} {98, 123523}

\bibitem[\protect\citeauthoryear{{Krochek} \& {Kovetz}}{{Krochek} \& {Kovetz}}{2022}]{2022PhRvD.105j3528K}
{Krochek} K.,  {Kovetz} E.~D.,  2022, \mn@doi [\prd] {10.1103/PhysRevD.105.103528}, \href {https://ui.adsabs.harvard.edu/abs/2022PhRvD.105j3528K} {105, 103528}

\bibitem[\protect\citeauthoryear{{Leung} et~al.,}{{Leung} et~al.}{2022}]{2022PhRvD.106d3017L}
{Leung} C.,  et~al., 2022, \mn@doi [\prd] {10.1103/PhysRevD.106.043017}, \href {https://ui.adsabs.harvard.edu/abs/2022PhRvD.106d3017L} {106, 043017}

\bibitem[\protect\citeauthoryear{{Liao}, {Zhang}, {Li}  \& {Gao}}{{Liao} et~al.}{2020}]{2020ApJ...896L..11L}
{Liao} K.,  {Zhang} S.~B.,  {Li} Z.,   {Gao} H.,  2020, \mn@doi [\apjl] {10.3847/2041-8213/ab963e}, \href {https://ui.adsabs.harvard.edu/abs/2020ApJ...896L..11L} {896, L11}

\bibitem[\protect\citeauthoryear{{Liao}, {Biesiada}  \& {Zhu}}{{Liao} et~al.}{2022}]{2022ChPhL..39k9801L}
{Liao} K.,  {Biesiada} M.,   {Zhu} Z.-H.,  2022, \mn@doi [Chinese Physics Letters] {10.1088/0256-307X/39/11/119801}, \href {https://ui.adsabs.harvard.edu/abs/2022ChPhL..39k9801L} {39, 119801}

\bibitem[\protect\citeauthoryear{Lorimer, Bailes, McLaughlin, Narkevic  \& Crawford}{Lorimer et~al.}{2007}]{Lorimer2007}
Lorimer D.~R.,  Bailes M.,  McLaughlin M.~A.,  Narkevic D.~J.,   Crawford F.,  2007, \mn@doi [Science] {10.1126/science.1147532}, 318, 777

\bibitem[\protect\citeauthoryear{{Masui} \& {Chime/Frb Collaboration}}{{Masui} \& {Chime/Frb Collaboration}}{2021}]{2021AAS...23832501M}
{Masui} K.,  {Chime/Frb Collaboration} 2021, in American Astronomical Society Meeting Abstracts. p. 325.01

\bibitem[\protect\citeauthoryear{{Mu{\~n}oz}, {Kovetz}, {Dai}  \& {Kamionkowski}}{{Mu{\~n}oz} et~al.}{2016}]{2016PhRvL.117i1301M}
{Mu{\~n}oz} J.~B.,  {Kovetz} E.~D.,  {Dai} L.,   {Kamionkowski} M.,  2016, \mn@doi [\prl] {10.1103/PhysRevLett.117.091301}, \href {https://ui.adsabs.harvard.edu/abs/2016PhRvL.117i1301M} {117, 091301}

\bibitem[\protect\citeauthoryear{{Oguri}}{{Oguri}}{2019}]{2019RPPh...82l6901O}
{Oguri} M.,  2019, \mn@doi [Reports on Progress in Physics] {10.1088/1361-6633/ab4fc5}, \href {https://ui.adsabs.harvard.edu/abs/2019RPPh...82l6901O} {82, 126901}

\bibitem[\protect\citeauthoryear{{Qiu}, {Keane}, {Bannister}, {James}  \& {Shannon}}{{Qiu} et~al.}{2023}]{2023MNRAS.523.5109Q}
{Qiu} H.,  {Keane} E.~F.,  {Bannister} K.~W.,  {James} C.~W.,   {Shannon} R.~M.,  2023, \mn@doi [\mnras] {10.1093/mnras/stad1740}, \href {https://ui.adsabs.harvard.edu/abs/2023MNRAS.523.5109Q} {523, 5109}

\bibitem[\protect\citeauthoryear{{Sammons}, {Macquart}, {Ekers}, {Shannon}, {Cho}, {Prochaska}, {Deller}  \& {Day}}{{Sammons} et~al.}{2020}]{2020ApJ...900..122S}
{Sammons} M.~W.,  {Macquart} J.-P.,  {Ekers} R.~D.,  {Shannon} R.~M.,  {Cho} H.,  {Prochaska} J.~X.,  {Deller} A.~T.,   {Day} C.~K.,  2020, \mn@doi [\apj] {10.3847/1538-4357/aba7bb}, \href {https://ui.adsabs.harvard.edu/abs/2020ApJ...900..122S} {900, 122}

\bibitem[\protect\citeauthoryear{{Tendulkar} et~al.,}{{Tendulkar} et~al.}{2017}]{Tendulkar2017}
{Tendulkar} S.~P.,  et~al., 2017, \mn@doi [\apjl] {10.3847/2041-8213/834/2/L7}, \href {https://ui.adsabs.harvard.edu/abs/2017ApJ...834L...7T} {834, L7}

\bibitem[\protect\citeauthoryear{{Treu}}{{Treu}}{2010}]{2010ARA&A..48...87T}
{Treu} T.,  2010, \mn@doi [\araa] {10.1146/annurev-astro-081309-130924}, \href {https://ui.adsabs.harvard.edu/abs/2010ARA&A..48...87T} {48, 87}

\bibitem[\protect\citeauthoryear{{Wang} \& {Wang}}{{Wang} \& {Wang}}{2018}]{2018A&A...614A..50W}
{Wang} Y.~K.,  {Wang} F.~Y.,  2018, \mn@doi [\aap] {10.1051/0004-6361/201731160}, \href {https://ui.adsabs.harvard.edu/abs/2018A&A...614A..50W} {614, A50}

\bibitem[\protect\citeauthoryear{{Wang}, {Jiang}, {Li}, {Ren}, {Tang}, {Zhou}, {Liang}  \& {Fan}}{{Wang} et~al.}{2021}]{2021ApJ...918L..34W}
{Wang} Y.,  {Jiang} L.-Y.,  {Li} C.-K.,  {Ren} J.,  {Tang} S.-P.,  {Zhou} Z.-M.,  {Liang} Y.-F.,   {Fan} Y.-Z.,  2021, \mn@doi [\apjl] {10.3847/2041-8213/ac1ff9}, \href {https://ui.adsabs.harvard.edu/abs/2021ApJ...918L..34W} {918, L34}

\bibitem[\protect\citeauthoryear{{Zhou}, {Li}, {Liao}, {Niu}, {Gao}, {Huang}, {Huang}  \& {Zhang}}{{Zhou} et~al.}{2022}]{2022ApJ...928..124Z}
{Zhou} H.,  {Li} Z.,  {Liao} K.,  {Niu} C.,  {Gao} H.,  {Huang} Z.,  {Huang} L.,   {Zhang} B.,  2022, \mn@doi [\apj] {10.3847/1538-4357/ac510d}, \href {https://ui.adsabs.harvard.edu/abs/2022ApJ...928..124Z} {928, 124}

\makeatother
\end{thebibliography}








\bsp	
\label{lastpage}
\end{document}